# A Real-Time Re-Scheduling Algorithm for Spacecraft Instrument Operations Optimization

**J. Pergoli[a,b]\*, T. Cesari[c,d,e]\*, M. Maestrini[b], P. Di Lizia[b], P. Luciano Losco[f]**

[a] *CLC Space GmbH, Alsbach- Hähnlein, Germany*
[b] *Politecnico di Milano, Milan, Italy*
[c] *University of Ottawa, Ottawa, Canada*
[d] *Università degli Studi di Milano, Milan, Italy*
[e] *Toulouse School of Economics, Toulouse, France*
[f] *SITAEL S.p.A, Rome, Italy*
\* Corresponding Authors

**Abstract**

The operation-planning of satellites, aimed at introducing a certain level of supervised automation during the execution of the operations, poses a great challenge to both designers and operators. From one side, the routine operations for an Earth Observation mission are predictable and typically repeatable; both these aspects are suitable for computerisation. On the other hand, not every non-nominal scenario can be anticipated and correctly formulated in terms of operations. Dealing with contingency presents risks which need to be addressed as early as possible, hopefully already during the operations preparation. It is also possible, however, to intervene at a later operational stage of the mission, optimising the tools already in use to support the operations execution. In this paper, having in mind the idea to improve existing processes in place at EUMETSAT, we present an algorithm able to reschedule the spacecraft's activities in case of anomaly. The main goal is to support the decision-making process while overcoming contingencies both avoiding overloading the spacecraft and planning engineers and ensuring the continuity of the mission, in particular giving the highest priority to the onboard computer memory size and data quality. We tested the method with the data of Sentinel-6, which carries the altimeter POSEIDON-4 operated by EUMETSAT, and the results are hereby presented.

**Acronyms/Abbreviations**

| | |
|---|---|
| CCSDS | Consultive Committee for Space Data Systems |
| EO | Earth Observation |
| ESA | European Space Agency |
| EUMETSAT | European Agency for the exploitation of Meteorological Satellites |
| FCT | Flight Control Team |
| FD | Flight Dynamic |
| FDIR | Failure Detection Isolation Recovery |
| GS | Ground Station |
| GUI | Graphic User Interface |
| HR | High Rate |
| LEO | Low Earth Orbit |
| LEOP | Launch and Early Orbit Phase |
| LR | Low Rate |
| LRM | Low Resolution Mode |
| MCC | Mission Control Center |
| MCS | Mission Control System |
| ML | Machine Learning |
| MP | Mission Planning |
| MPS | Mission Planning System |
| OBPS | On-Board Position Scheduler |
| SAR | Synthetic Aperture Radar |
| SAT-IOV | Satellite In-Orbit Validation |

# 1. Introduction

Space is a privileged perspective to observe the planet Earth, and for this reason data coming from EO satellites support a wide variety of applications. For example, remote sensing technologies that constantly monitor physical, chemical, and biological factors like the atmosphere composition or the sea level, serve the scientific community to understand the evolution of natural phenomena. In particular, given the undisputable fact that the human footprint is radically changing the whole natural system, it is paramount to find solutions to mitigate such impact for a more sustainable life on Earth. In essence, monitoring climate change is a critical factor for human survival.

One essential aspect to observe is the sea level, typically monitored by altimetry missions (see Table 1), which have become popular over the last few decades. An altimeter emits a microwave pulse and records part of its echo reflected by the surface. The time interval from the emission to the reception multiplied the speed of light at which electromagnetic waves travel gives the distance between the satellite and the Earth. By subtracting the altitude of the satellite, we obtain the surface height. These instruments have two main measurement modes: *Low-Resolution Mode* (LRM) and *Synthetic Aperture Radar* (SAR). The former provides measurements as the conventional radar altimeter, and the latter enhances the along-track resolution to measure the narrow portion of open water that LRM cannot achieves [1]. Figure 1 shows a typical example of altimetry measurements on different regions:

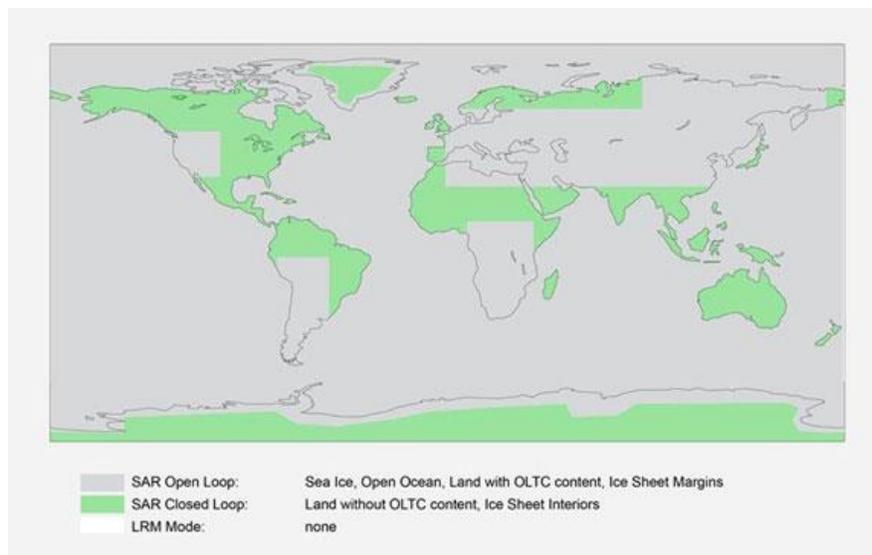

**Figure 1:** Examples of region where open and closed loop tracking may be appropriate (altimeter dependant)
(credit: ESA)

The *Mission Planning System* (MPS) is the interface between the user community, the space segment, and the ground segment. It optimizes the resources of the space segment to fulfil the user requirements. Traditionally, it is a tool which is extensively used during the routine operations, rather than during the early phases as *Launch Early Orbit Phase* (LEOP) or *Satellite In-Orbit Test* (Sat-IOV), for several reasons. First, because LEOP and SIOV are usually divided into phases which are critical as well as unique throughout the entire mission lifetime. These scenarios require more supervision and cannot always rely on cyclic automation. Secondly, because the personnel involved in routine operations can benefit from the experience gained during the early phases and can count on a more predictable operational cycle, precisely: the routine.

| Mission | Agency | Lifetime | Altitude [km] | Altimeter | Frequency | Repetitivity (day) |
|---|---|---|---|---|---|---|
| Cryosat-2 | ESA | April 2010 – Present | 720 | SIRAL | Ku band | 369 |
| HY-2 | China | August 2011 – Present | 971 | | Ku/C band | 14.168 |
| Saral | ISRO/CNES | February 2013 – Present | 800 | AltiKa | Ka band | 35 |
| Jason-3 | CNES/ NASA/ EUMETSAT/ NOAA | January 2016 – Present | 1336 | POSEIDON-3B | Ku/C band | 315 |
| Sentinel-3A | ESA/ EUMETSAT | February 2016 – Present | 814 | SRAL | Ku/C band | 27 |
| Sentinel-3B | ESA/ EUMETSAT | April 2018 – Present | 814 | SRAL | | 27 |
| Sentinel-6A | ESA/ EUMETSAT/ NASA/ NOAA | November 2020 – Present | 1336 | POSEIDON-4 | Ku/C band | 9.9965 |

**Table 1**: Recent altimetry missions.

Within the context of routine operations, there are at least three main aspects which shall be considered in order identify possible area of improvements:
- The traditional role of MPS within a ground segment. MPS leans on orbital events and activities from the flight control team and the ground systems [2]. The main point of MPS is allocating the desired tasks considering the constraints coming from the space segment (onboard memory size, spacecraft position), the ground segment (visibility opportunities), and the user requests. An efficient MPS requires appropriate software allowing quick planning and re-planning [3].
- The level of onboard automation of modern satellites. To yield better operations automation the onboard computer exploits the concept of position-based scheduling sub-service (OBPS) [4]: the capability to maintain an onboard position-based schedule of requests and to ensure the release of those requests at the associated orbit angles. At EUMETSAT, this concept is exploited to reduce the number of commands that the *Mission Control System* (MCS) prepares for the uplink every time there is a new weekly schedule.
- The basic operations concept of altimetry mission. In missions like altimetry, where the instrument mode changes accordingly to the surface that the spacecraft is flying over, the engineers upload different command sequences that are triggered according to the specific satellite orbit positioning.

In general, the existing MPS and the use of the OBPS mechanism simplify significantly the altimeter command chain, and the overall process grants a high level of robustness. However, a crucial problem appears every time that the system faces an unexpected event, such as ground station unavailability. In this case the planning engineer struggles to find a prompt solution that considers the altimetry mode changes, possibly exploits the OBPS service, minimizes the data loss, reduces the risk of the propagation of the anomaly (if applicable), maintains the operations as stable as possible and, eventually, supports efficiently the decision-making process of the management. Thus, having reactive MPS software could improve data quality and timeliness.

A possible way of addressing all these different needs is through the usage of *Machine Learning* algorithms. These methods have been proven to work effectively in an operational environment but require large training data sets [5]. A second possibility is by *Optimization algorithms*: they take out the need for the training, even though they might get 'only' *close-to-optimal* results [6]. In order to have an MPS ready for the use as soon as possible during the routine operations, when large training data is typically not available, *Optimisation Algorithms* have been preferred over *Machine Learning*. We take a path which in principle can be complex, having in mind that simplicity is the key for a successful operation centre.

We present an algorithm whose main objective is to optimize the onboard mass memory to dump data at the first possible opportunity. The algorithm considers all possible data rates of the instrument, the total memory occupied at a specific instance, and the different priorities between each mode to ensure that the satellite remains healthy, in line with the onboard FDIR mechanism. As the main case study, we consider the data produced by the Sentinel-6 spacecraft —operated by EUMETSAT— which carries an altimeter as the primary instrument.

The rest of the paper is organized as follows:

- Section 2 gives the mathematical settings and background.
- Section 3 presents the results and the future works.

- Section 4 discusses the benefits and limitations.

**2. Theory and calculations**

Usually, a nominal schedule for a Spacecraft is created with some margin in advance w.r.t the uplink, due to the numerous operations that need to be discussed thoroughly. The process grants robustness, but when unexpected events occur or there is not enough notice period, the identification of an optimal trade-off is not trivial at all. Sometimes, conservative choices to preserve the integrity of the mission at the cost of a lower data quality are preferred over solutions which would guarantee a higher data quality, however introducing risks for the entire programme.

To overcome this issue, the data acquisition is optimized in order to keep the quality as high as possible over critical targets reducing it everywhere else. The focus of the algorithm is to find the right trade-off to be still able to use only one dump to download data whilst guaranteeing the best data-rate possible.

We considered the case of an unexpected GS unavailability for the next $d_t$ orbits, thus forcing us to change the on-board instrument's operations to allow the on-board computer to download them at the first available pass without losing them and minimizing data degradation. We chose this scenario to show how reactive mission planning software could have improved the data quality and decreased the pressure on the operations teams to find a solution. We wanted also to underline that *'the easier'* is not always *'the better'*.

For all $n \in \mathbb{N}$, we denote the set $\{1, ..., n\}$ of the first $n$ positive integers by $[n]$. In what follows, we index *time* with a discrete variable $t \in \mathbb{N}$ representing the current orbit. More precisely, we denote the number of orbits of a cycle by $N \in \mathbb{N}$ and, for all $n \in \{0,1,2,...\}$ and $m \in [N]$, time $nN + m$ represents the $m$-th orbit of the $n$-th cycle.

Consider $K \in \mathbb{N}$ *modes*, represented here with the first $K$ integers $[K]$. Each mode $j \in [K]$ is associated with a *data rate* $\rho_j \geq 0$. We denote the free memory of the on-board computer at a time $t$ by $M_t$. The memory allocated for the calibrations during the $d_t$ missing orbits $\{t+1, ..., t+d_t\}$ after time $t$ is denoted by $C_{t,d_t}$. Each orbit $\tau$ is partitioned in a constant amount of $m_\tau$ of disjoint intervals $(u_{\tau,j})_{j=1,...,m_\tau}$ each with an assigned mode $\rho_{\tau,j} \in \{\rho_1, ..., \rho_K\}$. These intervals $u_{\tau,j}$ depend on the orbital geometric events.

The amount of memory that is normally allocated for $d_t$ orbits after a time $t$ is hence:

$$y_{t,d_t} := \sum_{\tau=t+1}^{t+d_t} \sum_{j=1}^{m_\tau} \rho_{\tau,j} u_{\tau,j}$$

*(1)*

Hence, the total memory that is required for $d_t$ orbits after a time $t$ is

$$M_t = y_{t,d_t} + C_{t,d_t} = \sum_{\tau=t+1}^{t+d_t} \sum_{j=1}^{m_\tau} \rho_{\tau,j} u_{\tau,j} + C_{t,d_t}$$

*(2)*

The memory, in nominal cases, is dumped every orbit over the designed ground stations. In the occasion of GS anomalies or unavailability, if the memory is not downloaded for a certain amount of time, the data start piling up and if at time $t$, data cannot be dumped for the next $d_t$ orbits and $y_{t,d_t} + C_{t,d_t} > M_t$; data loss will occur. Since calibrations of the instrument must be performed, $y_{t,d_t}$ needs to be decreased to prevent data loss. On the contrary, if $y_{t,d_t} + C_{t,d_t} < M_t$ nothing needs to be done.

Some modes must dwell for a fixed minimum time before switching to another one. For all orbits $\tau \in [N]$ and interval $j \in [m_\tau]$, let $\delta_{\tau,j} \geq 0$ be this time. We define a matrix of times $(\xi_{\tau,j})_{\tau \in [N], j \in [m_\tau]}$ that indicates how much time we want to spend in a specific mode, keeping in mind that $0 \leq \xi_{\tau,j} \leq u_{\tau,j} - \delta_{\tau,j}$; when $\xi_{\tau,j} = u_{\tau,j} - \delta_{\tau,j}$ we say that the mode associated with data-rate $\rho_{\tau,j}$ is a *prime mode*.

The idea is to divide each interval $u_{\tau,j}$ in two subsets containing, respectively, a $k_t$ and $(1 - k_t)$ fraction of $u_{\tau,j}$, for some constant $k_t \in [0,1]$ to be determined later.

We then use the nominal data-rate $\rho_{\tau,j}$ for the $k_t$ fraction and lowest possible positive data-rate $\rho_{\min} := \min\limits_{k\in[K], \rho_k > 0} \rho_k$ for the $(1 - k_t)$-fraction.

We want to maximize the time spent using $\rho_{\tau,j}$, so our objective function is:

$$f = \max_t (y_{t,d_t})$$

(3)

which is subjected to the following constraints:

$$y_{t,d_t} \leq M_t - C_{t,d_t}$$
$$\delta_{t,j} = \tilde{\delta}_{t,j}$$
$$\xi_{t,j} = \tilde{\xi}_{t,j}$$

(4)

Where $\tilde{\delta}_{t,j}$, $\tilde{\xi}_{t,j}$ are chosen by the operators. Considering also the division of each $u_{\tau,j}$, eq. (1) becomes:

$$\max_t(y_{t,d_t}) = \max_t \left( k_t \sum_{\tau=t+1}^{t+d_t} \sum_{j=1}^{m_\tau} (u_{\tau,j} - \delta_{\tau,j} - \xi_{\tau,j}) \rho_{\tau,j} + (1 - k_t)\rho_{\min} \sum_{\tau=t+1}^{t+d_t} \sum_{j=1}^{m_\tau} (u_{\tau,j} - \delta_{\tau,j} - \xi_{\tau,j}) \right)$$

(5)

The aim of the algorithm is to find the optimal value of $k_t$ which maximizes the amount of high-quality data. A direct verification shows that this is:

$$k_t := \frac{M_t - C_{t,d_t} - \sum_{\tau=t+1}^{t+d_t} \sum_{j=1}^{m_\tau} \xi_{\tau,j}\rho_{\tau,j} - \rho_{\min} \sum_{\tau=t+1}^{t+d_t} \sum_{j=1}^{m_\tau} (u_{\tau,j} - \delta_{\tau,j} - \xi_{\tau,j})}{\sum_{\tau=t+1}^{t+d_t} \sum_{j=1}^{m_\tau} (u_{\tau,j} - \delta_{\tau,j} - \xi_{\tau,j})\rho_{\tau,j} - \rho_{\min} \sum_{\tau=t+1}^{t+d_t} \sum_{j=1}^{m_\tau} (u_{\tau,j} - \delta_{\tau,j} - \xi_{\tau,j})}$$

(6)

This way, the nominal data-rate $\rho_{\tau,j}$ will still be used in the $u_{\tau,j}$ interval but for a shorter duration (the highest duration possible given the constraints), allowing the engineers to get images or measurements over the desired targets and reducing the data rate everywhere else.

### 2.1 Compressed Data Acquisition

The software needs to know the priority of each mode. Formally, it requires as input the matrices $(\delta_{\tau,j}), (\xi_{\tau,j})$, in addition to the usual intervals $(u_{\tau,j})$, corresponding data-rates $(\rho_{\tau,j})$, number $d_t$ of anomalous times after the present time $t$, available memory $M_t$, and amount of space $C_{t,d_t}$ needed for calibrations.

#### 2.1.1 Horizontal Compression

With the term horizontal compression, we indicate that the data acquisition is compressed by keeping the normal data rate $\rho_{\tau,j}$ only on a subset of the time window $u_{\tau,j}$. This means that we aim to keep quality as high as possible on the largest amount of time given the set or specific priorities $(\xi_{\tau,j})$.

**Algorithm 1**: CDA horizontal compression
Require: $0 \leq \xi_{\tau,j} \leq u_{\tau,j} - \delta_{\tau,j}$
**For** $t = 1, 2, \ldots$ **do**

**If** the spacecraft is notified with a GS anomaly message and no dumps are available for the next $d_t > 0$ orbits, **then**

    **If** $C_{t,d_t} > M_t$, **return fatal error** (the memory is not even sufficient to keep the instruments calibrated)
    **Else if** $y_{t,d_t} + C_{t,d_t} \leq M_t$, proceed nominally
    **Else if** $y_{t,d_t} + C_{t,d_t} > M_t$, keep nominal data-rates only on a fraction $k_t$ of times, as explained in **(6)**
    **End if**
**End if**
**End for**

*Table 2: Logic of CDA algorithm.*

## 3. Implementation & Results

To evaluate the performance of the method, we used as a test case the data produced by POSEIDON-4 [7], which is the altimeter on board the Copernicus Sentinel-6 spacecraft, a reference mission for high-precision ocean altimetry. The instrument allows the usage of both conventional *pulse-limited* (LRM) and *delay-Doppler* (SAR) to improve performance. In other words, the altimeter can swap between numerous modes, making the scheduling process a bit complex, especially during contingency. In our case, we experienced an outage of one of the GS for a few orbits, forcing us to rethink the spacecraft schedule during that time frame. First, we had to evaluate the amount of data, both in high and low quality, that the spacecraft would have saved during the anomaly's time. Secondly, we had to change the timing of the dumps to empty the memory and to be able to download all the archived memory during the first available pass without exceeding the visibility period. Last but not least, we deactivated high-quality data for the entire duration of the outage.

### 3.1. System Configuration

To allow the correct selection of each mode over each region of the Earth, the MPS receives every week the *geometric event file* from FDS [8]. This file contains the precise orbit position for each mode transition. MPS translates these transitions into command sequences considering users' requirements, and it ensures there aren't conflicts while commanding the instrument. MPS creates all the necessary commands using the *Drools Rules Language* [9], relying on the *Oracle* database. To streamline recurrent processes at EUMETSAT and to reproduce the scheduling S/W, the Mission Planning team developed an internal tool using Python 3.9. A web browser application was created using the Django framework [10] which emulates the actual operative system; besides, it eases the visual inspection of the MPS products. The reader can have an example of the *Graphic User Interface* in Figure 2:

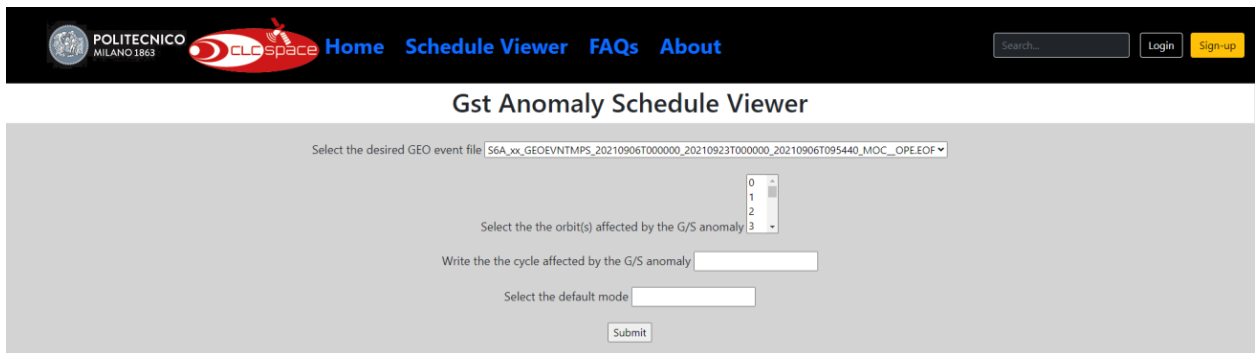

**Figure 2**: MPS GUI built with Django framework. From it the user can easily select specific FD files, the orbits where the anomaly occurs and also which default mode they would like to use instead of the nominal one in case of outages. In our case it is possible to select also the orbit cycle since the MPS for Sentinel-6 is based on the relative orbit number.

Through the GUI, the user can have a better overview of the impact of an issue impacting the spacecraft activities, for example, the amount of HR data that users can lose due to a malfunction of a GS. It is possible to select the time frame linked to the FD events, the number of orbits where the satellite cannot dump data, and the desired mode of the instrument that the engineers would like to have as default. Figure 3 shows the architecture of the framework:

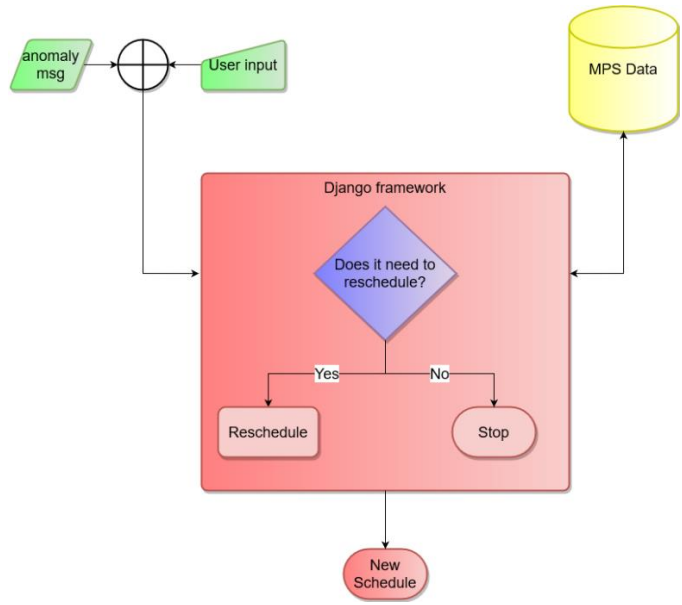

**Figure 3**: Architecture of the auxiliary MP S/W

### 3.2. Scenario Configuration

As anticipated, the data for the testing phase comes from POSEIDON-4, the altimeter on board the Sentinel-6 spacecraft. The instrument can be operated in three different main modes and undergoes various regular calibrations. The satellite flies in an LEO orbit with an altitude of 1336 km, inclination i of 66 deg, and 127 revolutions per cycle [11]. Each cycle lasts about 9.9 days. The computer on board can store up to 28 GB of high-rate data and 8 GB of low-rate data. The data dump occurs over two stations: Kiruna and Fairbanks.

### 3.3. Performance

During the commissioning phase, while characterizing the instrument's performance, we experienced an issue over one of the two antennas for five orbits. Hence, we had to record POS4 data only in LRM, except for calibrations. We decided to run our algorithm against this scenario to examine if we could have recorded high-rate data in the mass memory. The goal of the method is to get as much high-rate data as possible to dump the memory in one shot at the next available pass. The algorithm considers the different priorities among modes and calibration, e.g., the calibrations are mandatory, and some procedures are more critical than others. Figure 4 shows an example of the Gantt chart that our algorithm can produce during the contingency scenario:

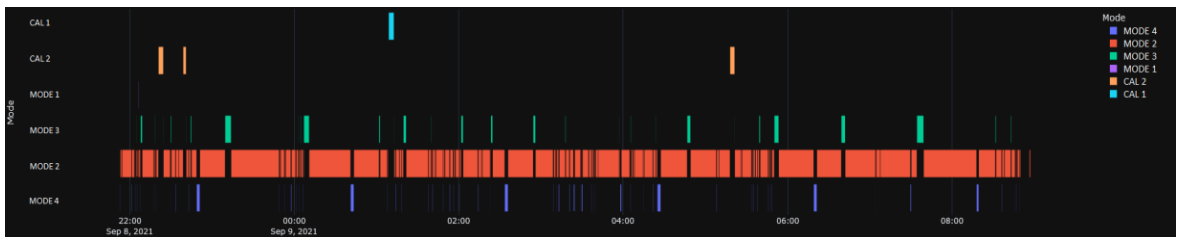

*(a)*

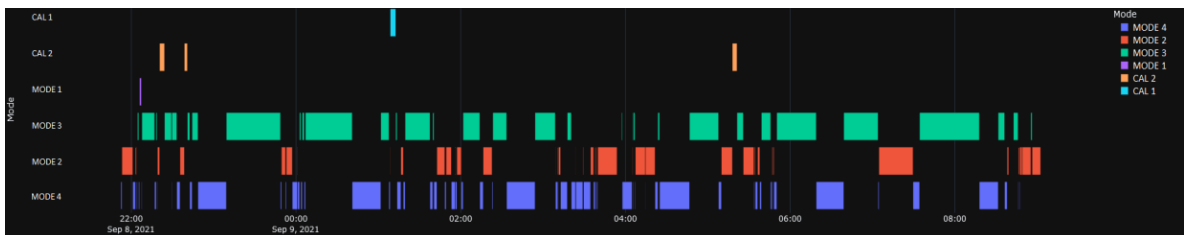

*(b)*

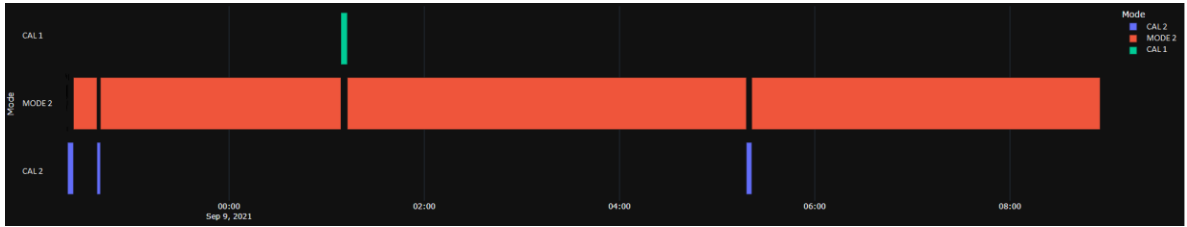

*(c)*

**Figure 4**: Comparison of the instrument activities during the outage scenario: picture (a) shows the Gantt chart of the rescheduling performed by our algorithm; (b) shows the Gantt chart during the same period if we had not experienced the outage and (c) shows the Gantt chart of the actual recorded data.

In Figure 4, we have compared three different cases: in the first, our algorithm generated the schedule considering an outage of 5 orbits and with Mode 1 as default mode; in the second, we created the instrument plan as if there were no issues; in the last case we used the schedule with the actual solution we decided to go for to register only LR data. The reader can see the benefit of our method: even if it is a small amount, we could have still recorded some HR over different regions. Table 3 sums up the numerical result for an easier inspection:

|  | **MODE 1 [%]** | **MODE 2 [%]** | **MODE 3 [%]** | **MODE 4 [%]** | **CAL 1 [%]** | **CAL 2 [%]** |
|---|---|---|---|---|---|---|
| **Manual re-scheduling** | 0 | 506 | 0 | 0 | 100 | 100 |
| **Automatic re-scheduling** | 14 | 474 | 12 | 18 | 100 | 100 |

*Table 3* shows the percentage of each instrument's mode in each schedule type compared against the operative one.

We decided to check what would have been the behavior of the algorithm in case of a shorter outage. Therefore, we performed a few tests where the issue lasted for five, three, and two orbits.

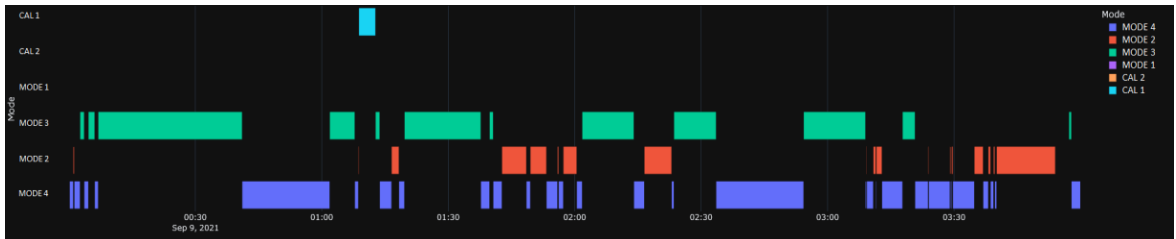

*(a)*

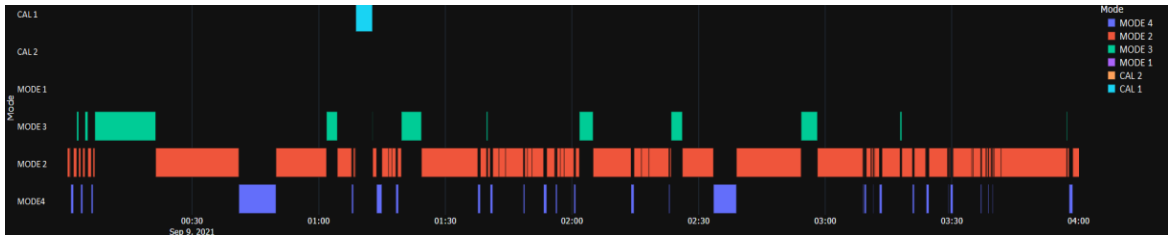

*(b)*

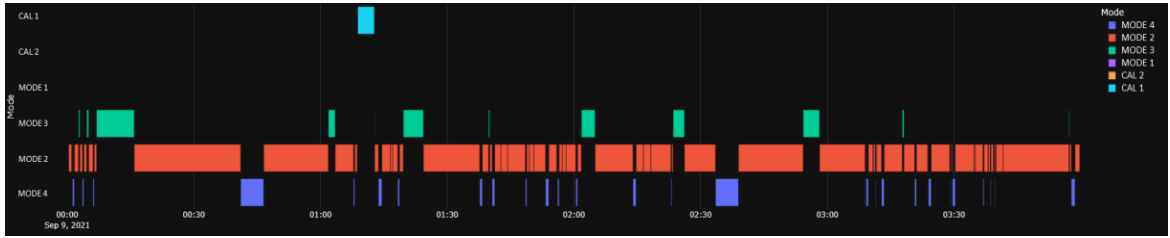

*(c)*

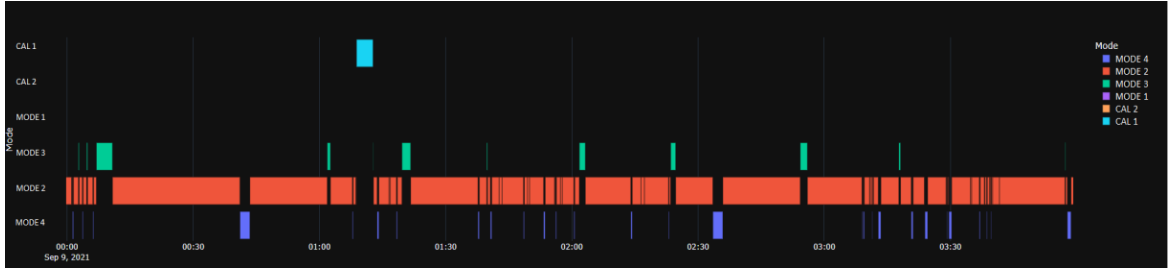

*(d)*

**Figure 5:** These four Gantt charts represent an outage of a GS progressively going from 2 orbits *()*, 3 orbits *()*, and 5 orbits *(d)* compared to nominal activities *(a)* in the time frame of 4 hours, from 00:00 UTC to 04:00 UTC on the 9[th] of September. It is clearly visible that the algorithm decreases the duration of some modes in order to fit the necessary ones.

The algorithm performed as expected: we set, again, LR data as the default mode, and we could see how all the others shrunk to accommodate the default.

*3.4. Future Works*

The method has been developed in a relative short amount of time and it is tailored on a specific altimeter mission, POSEIDON-4. For these reasons, we have at least two major areas for future improvements.

First of all, we shall improve the performance and the user experience at EUMETSAT. In particular:
- We already developed our method encapsulating the algorithm in a Django framework-based application to allow a friendly user experience. However, since the GUI can be expanded thanks to the numerous Python packages, we would like to add the definition of the various priorities to give the users better control of the main variables of the method.
- We need to optimize algorithm's performance by exploring other programming languages like *Rust* which offers different benefits like performance, community, and reliability.

Second, we need to test the algorithm on different missions and possibly on different ground segments, in order to make it as generic and flexible as possible.

## 4. Conclusion

In this paper we proposed a technique to automatically reschedule the ground operations via MPS, mainly when a contingency occurs. The main outcome of the algorithm is an array of operations, sorted by a pre-defined set of priorities and technical needs, which the mission engineer can use to implement the best possible sequence of commands during the recovery. This process drastically reduces the time that operators otherwise would spend in manual and repetitive tasks. The distinct advantage is to improve the quality of the job, as well as to support the decision-making process. The method is all but finished and few areas of improvement have been identified. However, the algorithm itself has been proven to be very useful while operating a LEO altimeter mission at EUMETSAT and we strongly believe that the results can be easily scaled to a generic EO mission. This method aims at improving an already existing MPS; however, it can be also considered as a valuable input for the design of future missions. Optimizing the onboard mass memory allocation, bearing in mind the users' requirements, could help mitigate other issues, like setting the correct timing to start and stop the data dump during the visibility period over a ground station, as studied by [12]. We believe that combining the different solutions could lead to a remarkable improvement in the way operators deal daily with spacecraft operations.

**Acknowledgements**

The authors gratefully acknowledge the support of the EUMETSAT Mission Planning team where Jonathan is working. This work started during Tommaso's Post-Doc at the Institut de Mathématiques de Toulouse, France, and benefited from the support of the project BOLD from the French national research agency (ANR). Tommaso Cesari gratefully acknowledges the support of IBM.